# UNCOVERING THE UNDERLYING PHYSICS OF DEGRADING SYSTEM BEHAVIOR THROUGH A DEEP NEURAL NETWORK FRAMEWORK: THE CASE OF RUL PROGNOSIS

Sergio Cofre-Martel[1], Enrique Lopez Droguett[1,2], Mohammad Modarres[1]

[1]University of Maryland, College Park, MD
[2]University of Chile, Santiago, Chile

## ABSTRACT

*Deep learning (DL) has become an essential tool in prognosis and health management (PHM), commonly used as a regression algorithm for the prognosis of a system's behavior. One particular metric of interest is the remaining useful life (RUL) estimated using monitoring sensor data. Most of these deep learning applications treat the algorithms as black-box functions, giving little to no control of the data interpretation. This becomes an issue if the models break the governing laws of physics or other natural sciences when no constraints are imposed. The latest research efforts have focused on applying complex DL models to achieve a low prediction error rather than studying how the models interpret the behavior of the data and the system itself. In this paper, we propose an open-box approach using a deep neural network framework to explore the physics of degradation through partial differential equations (PDEs). The framework has three stages, and it aims to discover a latent variable and corresponding PDE to represent the health state of the system. Models are trained as a supervised regression and designed to output the RUL as well as a latent variable map that can be used and interpreted as the system's health indicator.*

## 1. INTRODUCTION

In prognosis and health management (PHM), a prediction framework consists of four different technical processes [1]: Data acquisition, construction of health indicators, division of different health states, and health state prediction. The latter is commonly addressed through the estimation of the so-called remaining useful life (RUL) of the system, but it can also be related to other performance metrics, such as the State of Health (SOH) in Lithium-ion Batteries [2], or any other essential metrics for prognostics tasks in condition-based maintenance (CBM). Here, physics-based model (PBMs) [3] and data-driven approaches (DDAs) [4], [5] such as deep learning (DL) [6], and artificial intelligence [7], along with hybrid methods [8], have covered most research efforts in PHM for the last two decades.

The main difference between DDAs and PBMs techniques comes from the amount of information needed to adjust their models, as well as their capability to be adapted from one system to another. Physics-based models are known for their high accuracy due to the availability of a mathematical relationship that can describe the system's behavior. DDAs, on the other hand, are mostly presented as black-box algorithms that can extract information from a provided dataset. This gives DDA models the advantage of scalability and adaptability to any system. However, the black-box behavior of DDA comes also as a weakness, since these algorithms tend to be tested in benchmark datasets, and their results do not provide any interpretability. On the other hand, PBMs present the opposite behavior, where scalability and adaptability of the same model for different systems is rare, while model results are highly accurate and can be directly interpreted by the user. Since very few mathematical models are available for PBMs in prognostics [3], studies are normally limited to local crack propagation and corrosion. Thus, PBMs cannot be directly applied in complex systems or processes.

Hybrid methods have been proposed to combine PBMs and DDAs, and thus overcome their weaknesses and combine their strengths [8], [9]. Nevertheless, two essential challenges are yet to be addressed in DL models for prognostics: the use of time as an explicit variable, and the explicit relationship between the physics of the system and the input variables of the model. In the former, DL algorithms have been applied for prognostics purposes in a great variety of systems, such as lithium-ion batteries [10]–[13], rolling bearings [14]–[16], and turbofan engines [17]–[20], amongst others. However, these algorithms do not explicitly consider time as a variable in their calculations. Indeed, works through recurrent neural networks (RNN) and its long-short term memory (LSTM) variation [21]–[23], along with other hybrid deep learning architectures [9], use input data with time implicitly embedded through consecutive feature logs, which are then interpreted by the model. That is, the network is trained with a temporal sequence of data points to understand the given time scale represented in the data, rather than taking time as an explicit input for its direct interpretation. This can lead to inaccuracies in the models, since the network is given an additional task unrelated to the system's prognostics. Further, this can also induce overfitting and poor generalization, given that new data logs, previously unseen by the model, might have different temporal behavior in their log sequences.

Moreover, embedding the physics of degradation to a DL framework is a challenging task. This is mostly due to the lack of mathematical models that can represent these processes in complex systems. In this respect, reinforcement learning (RL) with neural networks (NN) have been recently implemented in the PHM context. Ballani et al. [24] addressed a sequential decision problem for optimal operation and maintenance with RL. Here, the Q-function is approximated by a NN per action. Mahmoodzadeh et al. [25] analyzed a pipeline corrosion physics-based testbench for CBM management using a novel RL framework. Results showed a 58% cost reduction compared to other methodologies, while improving the system's reliability. Both of these RL applications to PHM rely on the availability of an empirically-based mathematical model (i.e., crack propagation and corrosion, respectively) to describe the damage propagation or future behavior of the system. This presents a severe drawback since until today, acquiring PBM models represents one of the biggest challenges in PHM.

The latest advances in DL algorithms have shown that it is possible to embed PDEs to DL models. Raissi et al. [26] presented a physics-informed deep learning framework which allows solving PDEs using NN with only initial and boundary conditions, along with collocation points. They also showed how this idea could be implemented to discover PDEs embedded in the data by setting a penalty function that resembles a PDE-like behavior during the training of the neural network model. This opens the door to create a dynamic relationship between the input variables and prognostics metrics. Hence, in this paper, we present a deep neural network (DNN) framework for prognostics. The framework uses a latent variable representation to map the monitoring data, as well as time, to a one-dimensional space. The objective of the framework is to discover the underlying physics of degradation by modeling the RUL of a system, which is linked to the discovered latent variable, through a PDE-like penalization function. Once the model is trained, the latent variable can be used as a system health estimator, both quantitatively and qualitatively, through the RUL prediction and the latent variable mapping visualization. In other words, this framework resembles a PDE, where given initial feature values (i.e., initial conditions), the algorithm can estimate a RUL value through the PDE solution for a given time after the given initial conditions.

The remainder of this paper is structured as follows. Section 2 presents the background behind PDEs applied to DL. Section 3 illustrates the proposed DL framework, which is trained with the dataset presented in Section 4. The obtained results and their discussion are presented in Section 5. Section 6 outlines the main conclusions and remarks of this study.

## 2. PHYSICS-INFORMED DEEP LEARNING

Most applications of DL algorithms are used as black-box functions where the extraction of abstract relationships in the data is left for the machine to find. In this regard, efforts have recently gone into providing some interpretation and constraints to these techniques from a physics point of view. One such work comes from Raissi et al. [26], where a NN was implemented to model equations from fluid dynamics by only using initial and boundary conditions, as well as collocation points. In order to understand how these algorithms work, it is necessary to review how DL models work for function representation.

Deep learning is a branch of the machine learning discipline, where the main structure is deep neural networks, which are inspired as a functional representation of the human brain. Here, an input value is evaluated through sequential combinations of non-linear functions to yield the desired output value. Hence, one can represent the output $y$ of a NN as a function in the form of:

$$\hat{y} = f(X, W) \quad (1)$$

where, $f(X, W)$ is the NN, $X$ are the input values, and $W$ is a tensor of parameters called weights, which defines the function. Two key components compose a NN: layers and hidden units (also known as neurons). A layer is a non-linear function of an input value, commonly represented as:

$$h_i = \sigma(W_i^T h_{i-1} + b_i) \quad (2)$$

where $h_i$ is the hidden layer $i$, represented by its weight matrix $W_i$ and bias vector $b_i$. Notice that the relationship among $h_i$, $W_i$ and $X$ is a simple linear regression. Nevertheless, this linear expression is evaluated in a non-linear function $\sigma$, also referred to as activation function. The number of neurons or features from the previous layer and the number of neurons of the current layer give the dimensions of the weight matrix. As it can be observed in Equation 2, a layer takes as input the output of the previous layer, and it yields an output that goes on into the next hidden layer, and so on until the output layer is reached. For instance, Equation 3 shows a two-layer NN of input $X$, output $\hat{y}$, and activation function $\sigma$.

$$\hat{y} = \sigma(W_2^T \sigma(W_1^T X + b_1) + b_2) = f(X, W) \quad (3)$$

Thus, for a given dataset $(X, y)$, the parameters defining the NN in Equation 3 are trained (i.e., tuned or adjusted) to minimize the average of the squared errors, which is the so-called loss function described in Equation 4. Given a set of data points (often referred to as dataset), Equation 4 can be optimized using gradient descent [27] and backpropagation [28].

$$loss = \frac{1}{N} \sum_{i=1}^{N} (y_i - \hat{y}_i)^2 \quad (4)$$

On the other hand, PDEs are used to model the behavior of a function of interest based on the relationship between its partial derivatives with respect to its input variables. For instance, let $u(z, t)$ be a two-dimensional function of space and time. Then, a PDE for $u(z, t)$ can be represented as:

$$u_t = F(z, u, u_z, \dots) \quad (5)$$

where sub-indexes indicate partial derivatives of the function $u(z,t)$, e.g., $u_z = \partial u(z,t)/\partial z$. The right-hand side of Equation 5 is represented by a function $F$, that takes input variables related to the space variable $z$.

In their proposed methodology for physics-informed NNs, Raissi et al. [26] take advantage of automatic differentiation [31] to formulate a PDE-like penalization function. Considering the target value to estimate (e.g., velocity field, temperature, RUL) as the output of a NN expressed by $u(z,t)$. One can use automatic differentiation to calculate the exact derivative of the NN representing $u$ with respect to any of its input variables (e.g., $u_z, u_t$). This allows the creation of a PDE in the form of Equation 6, where $u_t$ is the time derivative of the output variable, and the function $F$ on the right-hand side is represented by a second NN, which takes the spatial variables and their corresponding $u$ derivatives as input. Equation 6 can be written as a cost function in terms of a function $f$ described in Equation 7. If $f$ is added to the training cost function of the DL model, it acts as a penalization term that binds the behavior of the parameters representing the NNs of $u(x,t)$ and $F$.

$$f := u_t - NN(x, u, u_x, \dots) = 0 \qquad (6)$$

Hence, the optimization cost function of the neural network can be written as the sum of the loss function in Equation 5, and the square of $f$:

$$Cost = \frac{1}{N}\sum_{i=1}^{N}(y_i - \hat{y}_i)^2 + \frac{1}{M}\sum_{j=1}^{M} f^2 \qquad (7)$$

where $M$ is the number of points to be tested in the PDE. These can be collocation points, initial conditions or boundary conditions.

## 3. PROPOSED FRAMEWORK

Most networks use all the data available to build very complex models, making it challenging to identify and control the effects of each input variable on the output of the model. A large number of input parameters or features might require a huge computational memory to process, but also can cause the model to overfit the training dataset. That is, the model might perform well when evaluated in the training samples, but it will fail to accurately evaluate new unseen values. This is known as poor generalization capability.

When training supervised deep learning models for RUL estimation, it is common to define a point at which the degradation process starts. This can be either a fixed time before failure [4], or when a certain performance variable surpasses a predefined threshold value [32]. Both of these approaches impose a strict constraint to the created RUL labels since it is assumed that the machinery understudy will continue to operate in the same condition until its failure. Hence, when applying DL algorithms to analyze these datasets and labels, the resulting trained models will be susceptible to errors when tested with new data. Nevertheless, if the deep network is trained to understand the dynamics of the degradation process, it would be more likely to improve its generalization capabilities. Indeed, adding a PDE-like penalization to the loss function of the model would create a relationship among the input features of the model and the derivatives of the output value with respect to its independent variables. This effect can be boosted if the framework is given time as an input feature, rather than implicitly extracting it from a sequence, since for metrics such as the RUL, the penalization function adds information on the degradation rate if temporal derivatives are considered.

Since there are no available equations that can directly map the health state of a complex system with its operational conditions, we propose a DNN framework to explore the degradation physics of a system through a latent variable representation. The supervised framework is aimed for PHM prediction tasks, where operational data is available from the monitoring of a system. To achieve this, the framework establishes a relationship between the latent variable and a prognosis output variable through a PDE-like penalization function. Figure 1 illustrates the proposed deep learning framework. The framework goal is to perform RUL predictions consisting of three stages that are represented by three different NNs. The first stage maps the operational conditions (OCs) and the time $t$ to a latent variable $x$. A second NN then takes both $t$ and $x$ to output the prognostics variable of interest, i.e., mainly the RUL of the system. A third NN is used to model the right-hand side of Equation 6, which models the time derivative of the RUL through a NN. This is the so-called dynamics of the PDE.

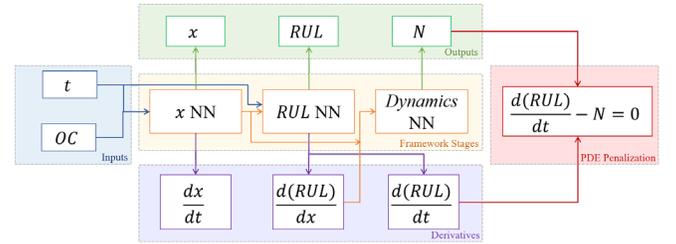

**FIGURE 1:** DEEP LEARNING ARCHITECTURE WITH LATENT VARIABLE AND PDE LOSS FUNCTION.

The NN for each stage of the proposed framework is structured as follows:
- ***x*-NN:** the network takes the OCs and time as input variables and it outputs the latent variable $x$. It is comprised of 5 hidden layers of 3 units each and one unit as output layer. Hyperbolic tangent (tanh) is used as the activation function.
- **RUL-NN:** it takes both the latent variable and time as input values and outputs the RUL of the system. It is comprised of 5 hidden layers of 10 units each. One output unit layer and tanh as the activation function.
- ***Dynamics*-NN:** it takes the time derivative of $x$, $dx/dt$, and $dRUL/dx$ as input values. It outputs a function $N$ that represents the dynamics of the system. This goes into the PDE-like penalization function. The network is comprised of 5 hidden layers of 10 units each. One output unit, and a rectified linear unit (ReLU) is used as the activation function.

Using automatic differentiation, we can take the time derivative from the first and second stage NN. These are then combined to form a PDE-like penalization term, as it is shown in Figure 1. The penalization includes the time derivative from the RUL, which is related to the Dynamics-NN in a PDE-like form. The penalization thus creates a dynamical relationship between the RUL and the latent variable $x$, which in turn is related to the initial operating conditions and the time at which the RUL is evaluated.

The framework addresses many of the drawbacks, as mentioned above, in DL methods for PHM. First and foremost, the network takes time as an input variable, along with the operational conditions of the system. The OCs are thus considered as initial conditions for a PDE, and the time corresponds to the point in the future at which it is desired to obtain the RUL value. In other words, for $t = 0$, the network behaves as most DL methods, that is, RUL is predicted based on the current OCs. Secondly, the use of a latent variable provides multiple advantages for both the training of the model, and the later interpretation of its results:

**Dimensionality reduction:** the usage of a latent variable facilitates the visualization of the results to make an informed decision based on the model's output. It also helps to capture and highlight important information related to the degradation process from the OCs.

**Input variables for Dynamic-NN:** the right-hand side function in Equation 6 could take every possible derivative from the input OC values. Thus, if all variables are considered in the Dynamics-NN, its input values would increase exponentially due to the number of derivatives that should be computed. This also would result in an increase in the training time.

**Eliminate redundancy and noise from the data**: due to the potentially high correlation among monitoring variables, it is common to observe that a system can be represented by a reduced variable space. This is the basic concept behind every data-driven approach for regression in PHM. Further, DNNs are known to remove noise levels in the input signals.

Note that out of the three stages, only the *RUL*-NN requires labels for the training process. Indeed, the latent variable $x$ comes as a secondary outcome from the backpropagation training of the *RUL*-NN. On the other hand, the Dynamics-network is trained solely from the penalization PDE term, which does not require any labels.

The proposed DL framework has two main outputs. It returns the RUL, as most DL PHM models do, and it also provides the trained latent variable $x$ along with its temporal derivative $dx/dt$, which can later be used for interpretation of the health state of the system when evaluating new unseen data. Since this latent variable was trained dynamically accounting for time and the RUL, its value should act as indicator of the health state of the system.

## 4. DATASET AND HARDWARE

The proposed model is tested using the benchmark dataset C-MAPSS due to the multiple research reports that have applied DL networks for its RUL estimation. Detailed description of this dataset and its processing can be found in [33], [34]. In this study, only the FD001 dataset was considered since the objective was to show the capability of the proposed framework to interpret the data through a PDE.

The dataset contains 27 sensor variables for 100 simulated engine runs, which are given a random initial degradation level. From this initial point, each engine is run until failure, recording the operational conditions for each cycle.

In order to train the proposed model, the original dataset needs some additional processing steps:
1. Select all data logs for one engine run, from its initial point until failure.
2. For each operational cycle, add a column with an integer time $t$ from 0 to 30.
3. Create a label for the above operation data and time, which corresponds to the RUL value at time $t$ since the initial point.

For instance, let us consider Engine 1, which contains a total of 192 log entries. If cycle number 100 is selected as the initial point, then for $t = 0$ its corresponding label is $RUL = 92$, then for $t = 1$ its label is $RUL = 91$, and so on until $t = 30$ is reached or until $RUL = 1$ (i.e., the engine fails). This process is repeated for each log entry of each engine, which increases the size of the original dataset from 20,631 to 593,061 points.

Models are trained on Python 3.6 with the use of Tensorflow 2.0 and Keras. Windows is used as operating system. The computer hardware consists on an Intel i7-9700k CPU, 32 GB of RAM memory, and a 24 GB Titan RTX GPU. The average training time in this machine is 140 seconds, while the evaluation time for new data entries is 0.01 seconds.

## 5. RESULTS AND DISCUSSION

For the training of the model, 25% of the training set is randomly selected to be used as a validation dataset. The model is then trained on 444,795 samples and subsequently validated on the 148,266 unseen samples. The training process is done using the NAdam optimizer [35]. Figure 2 illustrates the training and validation loss throughout the training process. It can be observed that the model does not present overfitting nor underfitting since both curves present an identical behavior. Also, they converge to the same loss value. Thus, the model presents good generalization capabilities. This can be attributed to the PDE penalization function added to the model. Indeed, the dynamical relationship built between the latent variable and the RUL as well as the inclusion of the time dimension, give the model more information. In turn, this makes the model capable of yielding more consistent predictions.

After 10 different training processes, each with random initial weights for each NN, the framework averages a root mean square error (RMSE) value of 17.8 cycles. Figure 3 shows the RUL predictions compared with their target value for the 100 engines provided in the FD001 test set. Although this RMSE value is not as low as other far more complex architectures, it is still in the acceptable range [20]. Moreover, the most important output of our proposed framework comes from the latent variable representation, which is illustrated in Figure 4.

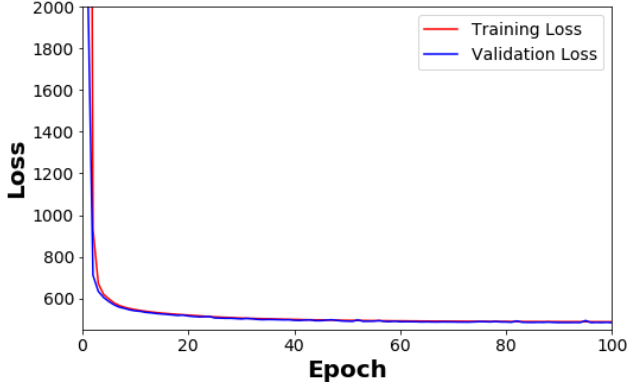

**FIGURE 2:** TRAINING AND VALIDATION LOSS.

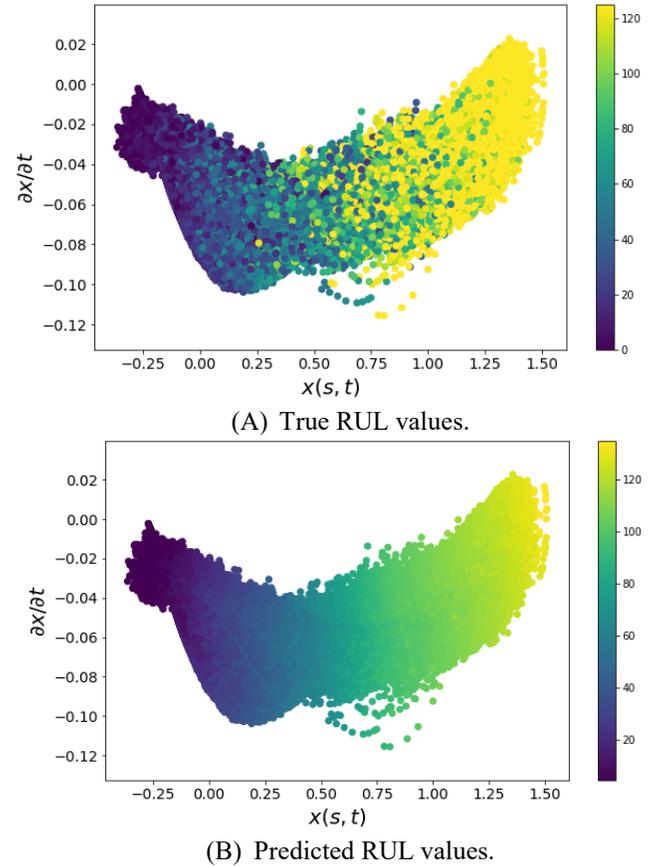

(A) True RUL values.

(B) Predicted RUL values.

**FIGURE 4:** RUL VALUES MAPPED TO LATENT VARIABLE AND ITS TEMPORAL DERIVATIVE FOR (A) TRUE RUL LABELS (B) PREDICTED RUL VALUES.

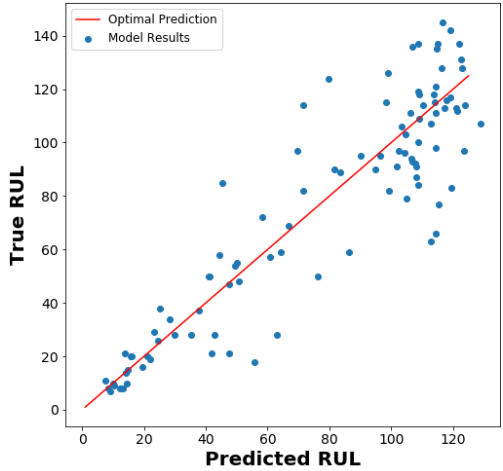

**FIGURE 3:** TRUE VS PREDICTED RUL VALUES.

Indeed, Figure 4 shows the mapping of the predicted RUL to the corresponding estimated latent variable given the training input data. Here, the latent variable $x$ and its time derivative $dx/dt$ are plotted with the RUL value represented as a color map. Figure 4a shows the RUL values corresponding to the true training labels, while Figure 4b shows the RUL values yielded by the model. We can see that the proposed framework smooths the latent variable space, creating a continuous relationship between the operational conditions and the RUL. Figure 4b can be used to relate the health state of the system with the latent variable and its time derivative. In turn, this gives useful information as a CBM assessment tool. For instance, by establishing a desired RUL threshold value, Figure 4b could be used as an inspection and maintenance scheduling tool. Here, a classifier can be created to separate the data between healthy and degraded state according to its $x$ and $dx/dt$ values. An online implementation of the model, along with the classifier, would allow a real-time evaluation of the operational conditions of the system. This classifier can be further complemented with the remaining stages of the framework presented in Figure 1, i.e., the PDE dynamics $N$ and the $RUL$ estimation. These additional outputs provide information on the system and can be used to create new metrics, rather than just base the results on a RUL value. Thus, opening the opportunity to make a better-informed decision on the maintenance scheduling of the system. Note that given its nature, the Dynamics-NN will be strongly related to the physics degradation of the system under study, and thus the impact of its results will vary for different applications.

Since the model takes time as an input variable, a second application could involve creating different maps for multiple future time horizons given a set of initial operational conditions. As an example, for a given a set of operational conditions $OC^*_{t=0}$, the proposed framework allows the estimation of the RUL for different time horizons, e.g., $t = 1$ and $t = 2$. These RUL values and their corresponding latent variable $x(OC^*_{t=0}, t)$ can thus be allocated in the RUL map presented in Figure 4b for any given time in the future. Hence, taking time as an input variable allows the model to trace, compare, and analyze the temporal evolution of $x$ and its RUL estimation based on the current operational conditions. Another possibility is to estimate multiple RUL values for a given time horizon $t^*$ when OC data is available for different points in time. For instance, for $t^* = 30$, if OC are available for times $t = [0,29]$, i.e., $[OC_{t=0}, OC_{t=1}, ..., OC_{t=29}]$, we can obtain 30 different RUL estimations for the same time horizon. These examples show how the inclusion of time and the degradation physics represented through the derivatives within the model are the main advantage of the proposed framework since it gives flexibility to the results that can be produced.

Figure 5 illustrates the RUL predictions when evaluating the model with the test set. It can be observed that the latent variable follows the same pattern as Figure 4b. This corroborates the generalization capabilities of the model shown in Figure 2. This also supports the argument that the model can be used as a health assessment tool, since it yields consistent results with unseen data points. Note also that the resulting latent variable will depend on the initial weights of the network. In this case, these weights are initialized with a standard normal distribution. Hence, depending on the initial weights, Figure 4 can present different shapes. Nevertheless, once the model yields satisfactory results, these initial weight values can be fixed with the purpose of obtaining repeatable results if the model is to be trained with new data.

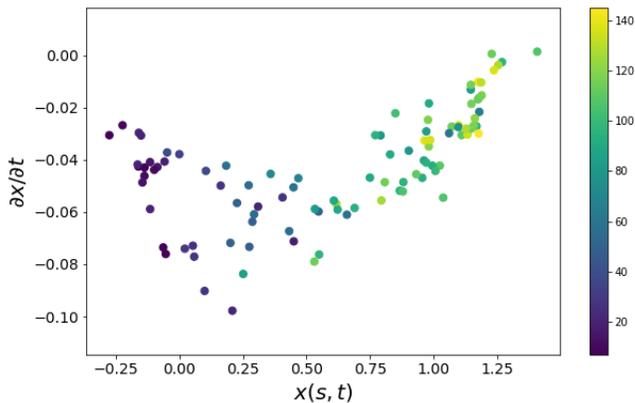

**FIGURE 5:** RUL MAPPING FOR THE TEST SET WITH RESPECT TO THE LATENT VARIABLE AND ITS TEMPORAL DERIVATIVE.

## 6. CONCLUSIONS

In this paper we have presented a DL framework for RUL estimation which explores the physics of degradation through a latent variable and a PDE-like penalization function. The framework uses time as an input value, while the discovered latent variable allows the interpretation of the operational conditions from an engineering point of view. The framework opens many doors to the application of these algorithms to real complex systems, especially on maintenance and preventive assessments.